\documentclass[twocolumn,aps,english,superscriptaddress]{revtex4-1}
\usepackage[T1]{fontenc}
\usepackage[latin9]{inputenc}
\usepackage{amsmath}
\usepackage{amssymb}
\usepackage{cancel}
\usepackage{graphicx}
\usepackage{esint}
\usepackage{color}
\makeatletter

\@ifundefined{textcolor}{}{%
 \definecolor{BLACK}{gray}{0}
 \definecolor{WHITE}{gray}{1}
 \definecolor{RED}{rgb}{1,0,0}
 \definecolor{GREEN}{rgb}{0,1,0}
 \definecolor{BLUE}{rgb}{0,0,1}		
 \definecolor{CYAN}{cmyk}{1,0,0,0}
 \definecolor{MAGENTA}{cmyk}{0,1,0,0}
 \definecolor{YELLOW}{cmyk}{0,0,1,0}
}

\usepackage{dcolumn}
\usepackage{bm}
\usepackage{enumerate}

\newcommand{\Ham}{\hat{H}}

\newcommand{\OPc}[2]{\hat{#1}_{#2}^{\dag}}
\newcommand{\OP}[2]{\hat{#1}_{#2}^{\vphantom{\dag}}}
\newcommand{\CD}[1]{\OPc{c}{#1}}
\newcommand{\C}[1]{\OP{c}{#1}}

\newcommand{\hc}{\textrm{h.c.}}
\newcommand{\E}{\epsilon}

\newcommand{\expect}[1]{\left<#1\right>}

\renewcommand{\k}{\mathbf{k}}
\newcommand{\q}{\mathbf{q}}

\newcommand{\ND}[1]{\hat{n}_{#1}}

\newcommand{\captiontitle}[1]{{\bf #1}}

\newcommand{\DD}[1]{\OPc{d}{#1}}
\newcommand{\D}[1]{\OP{d}{#1}}

\newcommand{\blochVec}[3]{{\left[ #1,~ #2,~ #3 \right]^\top}}

\DeclareMathOperator{\tr}{\textrm{Tr}}

\usepackage{babel}
\addto\captionsenglish{}
\renewcommand{\thefigure}{\@arabic\c@figure}

\makeatother

\begin{document}
\title{Universal Optical Control of Chiral Superconductors and Majorana Modes}
\date{\today}
\author{M. Claassen}
\affiliation{Center for Computational Quantum Physics, Simons Foundation Flatiron Institute, New York, NY 10010 USA}
\author{D. M. Kennes}
\affiliation{Dahlem Center for Complex Quantum Systems and Fachbereich Physik, Freie Universit\"{a}t Berlin, 14195 Berlin, Germany}
\author{M. Zingl}
\affiliation{Center for Computational Quantum Physics, Simons Foundation Flatiron Institute, New York, NY 10010 USA}
\author{M. A. Sentef}
\affiliation{Max Planck Institute for the Structure and Dynamics of Matter, Center for Free Electron Laser Science, 22761 Hamburg, Germany}
\author{A. Rubio}
\affiliation{Center for Computational Quantum Physics, Simons Foundation Flatiron Institute, New York, NY 10010 USA}
\affiliation{Max Planck Institute for the Structure and Dynamics of Matter, Center for Free Electron Laser Science, 22761 Hamburg, Germany}

\maketitle



\textbf{Chiral superconductors are a novel class of unconventional superconductors that host topologically protected chiral Majorana fermions at interfaces and domain walls \cite{readgreen2000,schnyder08,KallinProgPhys2015}, elusive quasiparticles \cite{ElliottRMP2015,BeenakkerAnnuRevCondMat2013,AliceaProgPhys2012} 
	that could serve as a platform for topological quantum computing \cite{nayak07}.
	Here we show that, in analogy to a qubit, the out-of-equilibrium superconducting state in such materials can be described by a Bloch vector and controlled on ultrafast time scales. The all-optical control mechanism is universal, permitting arbitrary rotations of the order parameter, and can induce a dynamical change of handedness of the condensate.	It relies on transient breaking of crystal symmetries via choice of pulse polarization to enable arbitrary rotations of the Bloch vector. The mechanism extends to ultrafast time scales, and importantly the engineered state persists after the pump is switched off. We demonstrate that these phenomena should appear in graphene \cite{blackschaffer06,nandkishore11,kiesel11} or magic-angle twisted bilayer graphene (TBG) \cite{cao2018,mcdonaldPNAS2011,liu18,kennes18}, as well as Sr$_2$RuO$_4$ \cite{luke98,mackenzie03}, as candidate chiral $d+id$ and $p+ip$ superconductors, respectively. Furthermore, we show that chiral superconductivity can be detected in time-resolved pump-probe measurements. This paves the way towards a robust mechanism for ultrafast control and measurement of chirally-ordered phases and Majorana modes.}

\begin{figure*}[t]
\centering
\includegraphics[width=\textwidth]{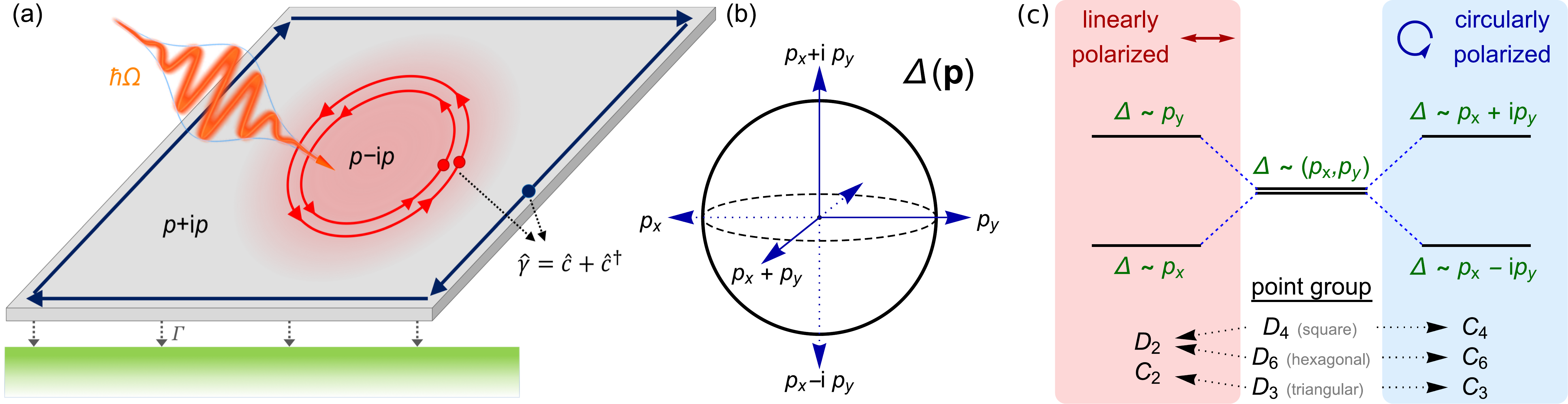}
\caption{\captiontitle{Driven chiral superconductors and dynamical symmetry breaking of two-dimensional order parameters.} (a) Schematic of a photo-induced domain with opposite chirality in a $p+ip$ superconductor. Induced domain walls host two chiral Majorana edge modes (red arrows in a), and persist after the pump is off, while the system equilibrates locally by dissipating energy with rate $\Gamma$. (b) Bloch vector representation of the non-equilibrium pairing function of a chiral superconductor. Optical pumping induces arbitrary rotations of the order parameter vector via choice of polarization. (c) Irradiation with linearly-polarized light breaks the discrete lattice rotation symmetry, lifting the degeneracy between $p_x, p_y$ order parameters. Circularly-polarized light breaks both the coplanar reflection symmetries and time-reversal, splitting the degeneracy into conjugate chiral representations $p_x\pm ip_y$.}
\label{fig:schematics}
\end{figure*}

The search for Majorana fermions in condensed matter has recently generated much excitement \cite{ElliottRMP2015,BeenakkerAnnuRevCondMat2013,AliceaProgPhys2012}, in particular due to their potential use in topological quantum computing \cite{nayak07}. A particularly intriguing realization can be found in quasi-two-dimensional chiral superconductors \cite{KallinProgPhys2015}, for which the phase of the order parameter $\Delta(\k)$ winds in either clockwise or counter-clockwise fashion about the Fermi surface, thereby spontaneously breaking time-reversal symmetry (TRS). Imbued with a non-trivial $\mathbb{Z}$ bulk topology \cite{schnyder08}, such materials host topologically-protected chiral Majorana modes at their boundaries as well as vortices with non-Abelian statistics \cite{AliceaProgPhys2012,KallinProgPhys2015,ElliottRMP2015}. 

Over the past years, experimental evidence for the requisite TRS breaking and spin-triplet pairing has accumulated for Sr$_2$RuO$_4$ \cite{luke98,mackenzie03} 
and UPt$_3$ \cite{luke93,Joynt02}, rendering these materials candidate chiral $p+ip$ or $f+if$ superconductors. Moreover the recent discovery of superconductivity in magic-angle twisted bilayer graphene (TBG) \cite{cao2018} has spurred proposals of exotic $d_{x^2-y^2}+id_{xy}$ singlet pairing \cite{liu18,kennes18}, analogous to predictions for heavily-doped graphene monolayers \cite{blackschaffer06,nandkishore11,kiesel11}. However, a central follow-up question is how chiral condensates and associated Majorana modes can be probed and controlled.

Here we show that optical pumping permits universal control of the order parameter and handedness of chiral superconductors, and can serve as direct evidence of their chiral nature in time-resolved pump-probe experiments via tracking transient signatures of the nonequilibrium electronic order \cite{mitrano2016c60,sentef16}. Central to the robustness of our predictions, the underlying effect works on ultrafast time scales, persists after the pump is off, and relies solely on symmetry arguments, hence making it applicable to any material described by a chiral order parameter. An immediate consequence is the possibility to optically define domains with a particular handedness of the chiral order parameter with chiral Majorana modes at their boundaries, depicted schematically in Fig. \ref{fig:schematics}(a).

In unconventional superconductors, TRS can be spontaneously broken if the nodal pairing function is degenerate due to crystal symmetry \cite{sigrist91}. The simplest manifestation is triplet $p$-wave pairing in a tetragonal crystal, where $p_x$ and $p_y$ nodal gap functions span a two-dimensional irreducible representation (irrep) $E_u$.  The degeneracy derives from joint $C_4$ rotation and $\sigma_v, \sigma_d$ mirror symmetries. Below the superconducting critical temperature $T_C$ it is energetically favorable to form a $p_x + i p_y$ condensate and spontaneously break TRS and parity in order to open a gap across the entire Fermi surface. As the choice of handedness is arbitrary, such a chiral pairing function is degenerate with its time-reversed Kramers partner $p_x - i p_y$. Similarly, in hexagonal $D_{6}$ or $D_{3}$ systems such as graphene or magic-angle TBG with $C_6$ or $C_3$ rotation and in-plane mirror or out-of-plane two-fold rotation symmetries, $d_{x^2-y^2}$, $d_{xy}$ are degenerate, spanning the $E_{2}$ or $E$ two-dimensional irrep, such that a chiral $d_{x^2-y^2} \pm i d_{xy}$ state becomes energetically favorable. In equilibrium or steady state, the symmetries underpinning this degeneracy can be selectively broken via uniaxial pressure \cite{sigrist02,Hicks283}, high-frequency Floquet pumping \cite{dehghani17} or magnetic fields \cite{agterberg98}, to cool or quench from the normal state to a superconducting state with a particular nodal or chiral order.

\begin{figure*}[t]
	\centering
	\includegraphics[width=\textwidth]{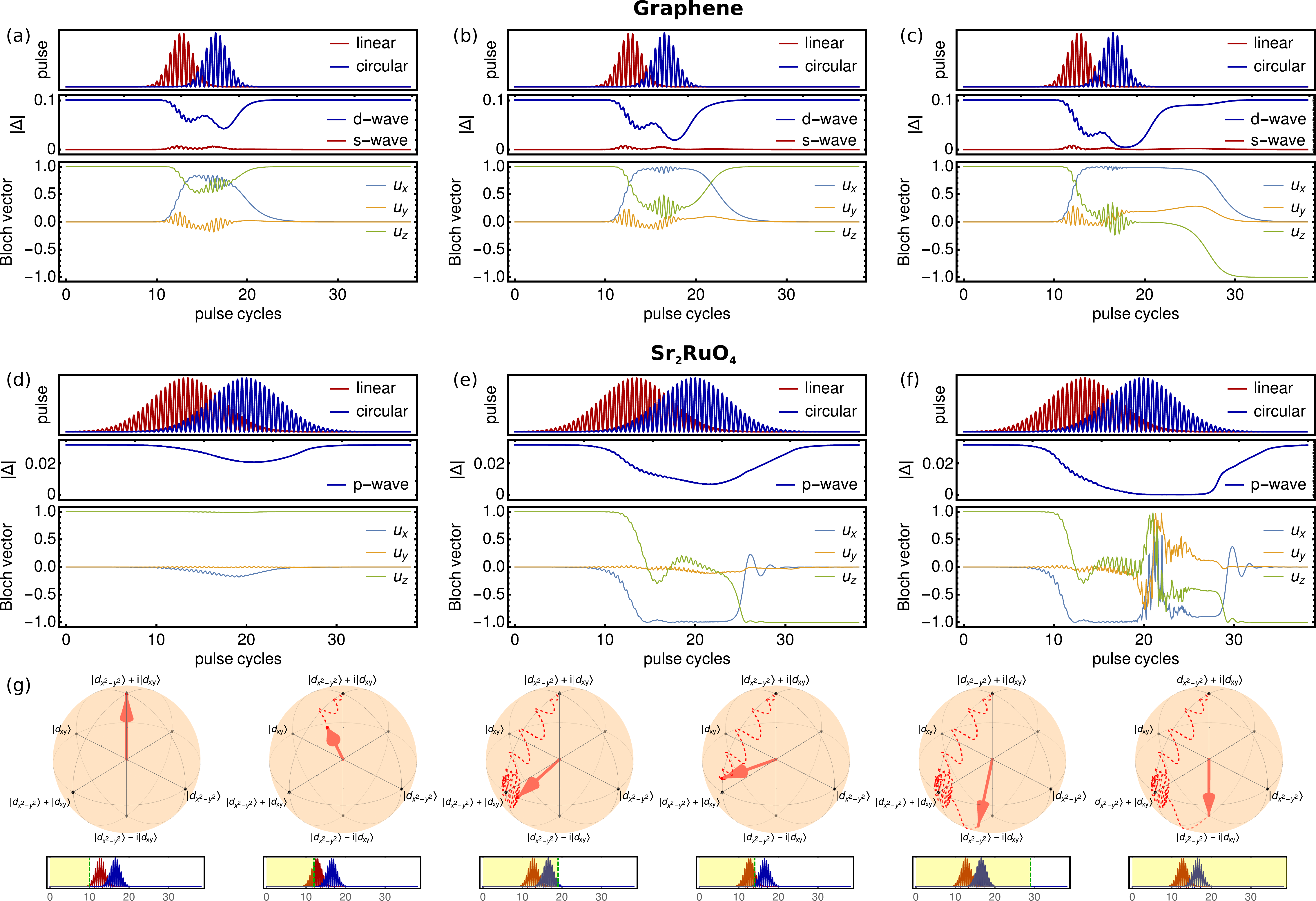}
	\caption{\captiontitle{Ultrafast Control of Chirality of the Superconducting Order Parameter.} Evolution of the Bloch vector and magnitude of the superconducting gap, for an ultrafast two-pulse sequence with linear and circular polarization, as a function of pump strength. (a)-(c) depicts the order parameter dynamics starting from a $d_{x^2-y^2} + i d_{xy}$ state in graphene, for dimensionless pump strengths $A=0.25, 0.35, 0.4$, from left to right, for $\omega = 0.8 \Delta_{\textrm{eq}}$ and a short pulse envelope. The linearly-polarized pulse rotates the Bloch vector from $[0,~0,~1]^\top$ to a nodal state with $[0,~1,~0]$; the second circularly-polarized pulse selects a state with opposite chirality $[0,~0,~-1]^\top$ beyond a critical pump strength. Weak coupling to extended $s$-wave pairing is allowed via symmetry but negligible in the parameter regimes considered here. (d)-(f) depicts the optical reversal of chirality from $p_x + ip_y$ to $p_x - ip_y$ in an effective three-band model for Sr$_2$RuO$_4$, for $A=0.1,0.2,0.3$ from left to right. (g) Switching of the order parameter Bloch vector as in (c) but represented on the Bloch sphere (see supplemental information for the corresponding movie). 
	}
	\label{fig:UltraFast}
\end{figure*}

The principal observation that constitutes the starting point for this work is that a chiral order parameter pushed out of equilibrium via irradiation will cease to be constrained to the equilibrium free-energy minima $p+ip$ or $p-ip$ ($d+ id$ or $d-id$), and instead evolve within the full two-dimensional subspace spanned by the corresponding irrep. Generically, a spatially homogeneous superconducting gap function within a two-dimensional irrep can be written as
$
	\boldsymbol{\Delta}(\k) = \Delta \left[ \cos\left(\frac{\theta}{2}\right) \hat{\boldsymbol{\eta}}^{(1)}_\k + e^{i\phi} \sin\left(\frac{\theta}{2}\right) \hat{\boldsymbol{\eta}}^{(2)}_\k  \label{eq:BlochVector} \right]
$.
Here, $\Delta$ is the gap amplitude and $\hat{\boldsymbol{\eta}}^{(1,2)}_\k$ are explicit form factors, for instance a chiral representation $\eta_\k^{(1,2)} = \sin(k_x) \pm i \sin(k_y)$ for nearest-neighbor $p$-wave pairing on the square lattice. One immediately sees that the out-of-equilibrium order parameter can be represented via a Bloch vector in direct analogy to a qubit [Fig. \ref{fig:schematics}(b)], with 
$
	\mathbf{u} = \left[ \sin(\theta) \cos(\phi),~ \sin(\theta) \sin(\phi),~ \cos(\theta) \right]^\top
$.
We choose $\mathbf{u} = \left[0,~0,~ \pm 1\right]^\top$ to denote chiral $p_x \pm i p_y$ or $d_{x^2-y^2}\pm id_{xy}$ pairing, with nodal order parameters lying on the Bloch sphere equator. 

An optical pulse can now selectively reduce the symmetry group via choice of light polarization and induce controlled dynamics of the Bloch vector. A linearly-polarized laser pulse necessarily breaks $C_3$, $C_4$ or $C_6$ discrete lattice rotation symmetries, while preserving mirror symmetries if the polarization vector is collinear with a crystallographic direction. The point group reduces from $D_{4}$, $D_{6}$ or $D_{3}$ to $D_{2}$ or $C_{2}$, and the two-dimensional $E$ or $E_2$ irrep splits into one-dimensional irreps $B_2$, $B_3$ or $B$. This breaks the degeneracy between $p_x$, $p_y$ or $d_{x^2-y^2}$, $d_{xy}$ pairing functions [Fig.~\ref{fig:schematics}(c), left panel], and can be thought of as transiently changing their respective condensation temperatures. 

If the system is initially in a state $p_x + ip_y$ with $\mathbf{u} = \left[ 0,~ 0,~ 1 \right]^\top$, a linearly-polarized pulse ensures that the chiral form factors $p_x \pm i p_y$ cease to represent eigenstates of the pairing problem for the duration of the pulse due to broken rotation symmetry. Hence, the Bloch vector $\mathbf{u}$ starts to oscillate between nodal $p_x$ and $p_y$ pairing with $\mathbf{u} = \left[ 1,~ 0,~ 0 \right]^\top$, $\left[ -1,~ 0,~ 0 \right]^\top$, respectively.

Conversely, a circularly-polarized pump pulse with collinear sample and polarization planes breaks the in-plane mirror symmetries $\sigma_v, \sigma_d$ as well as TRS while retaining discrete rotation and spin-inversion symmetry. The point group reduces from $D_{4}$ to $C_{4}$ (or from $D_{3}, D_{6}$ to $C_{3}, C_{6}$ in triangular or hexagonal lattices), and the two-dimensional $E$ irrep splits into chiral complex-conjugate one-dimensional irreps [Fig.~\ref{fig:schematics}~(c)]. In this case, the degeneracy of order parameters is lifted in favor of two chiral one-dimensional representations $p_x + ip_y$ and $p_x - ip_y$. Analogously, the two-dimensional irrep for $d$-wave pairing in hexagonal systems reduces to one-dimensional complex-conjugate representations $d_{x^2-y^2} \pm i d_{xy}$.

Starting from $p_x + ip_y$, a circularly-polarized pulse will now merely induce oscillations in the gap amplitude $\Delta$ while leaving the Bloch vector and chirality of the pairing function inert. Instead, if the superconductor is prepared in a nodal state, $p_x$ and $p_y$ cease to represent eigenstates of the pairing problem due to the circularly polarized laser and the Bloch vector will oscillate between chiral $p_x + ip_y$ and $p_x - ip_y$ states with $\left[ 0,~ 0,~ \pm1 \right]^\top$. 

These considerations immediately suggest that arbitrary (universal) rotations of the order parameter can be induced by sequences of linearly, circularly or elliptically polarized pulses. Intriguingly, starting from a chiral state, an appropriately-shaped \textit{linearly-polarized} pulse can perform a $\pi$-rotation of the Bloch vector and induce a domain with reversed chirality and topologically-protected Majorana modes along the domain wall boundaries [Fig. \ref{fig:schematics}(a)]. Once the pulse is switched off, the rotated order parameter remains a steady state of the equilibrium Hamiltonian.

While the ideal pulse shape and duration for a linearly-polarized $\pi$ pulse generically depends sensitively on materials details, the protocol for reversal of chirality is not unique. Instead, the symmetry considerations above suggest an ultrafast two-pulse sequence of linearly- and circularly-polarized pulses as a simple robust choice. Here, the linearly-polarized pulse liberates the order parameter from its free energy minimum, rotating the Bloch vector toward a nodal state, while the handedness of the second circularly-polarized pulse subsequently fixes the final handedness of the chiral domain.

We now apply the above generic mechanism to $d+id$ superconductivity in graphene \cite{blackschaffer06,nandkishore11,kiesel11,liu18,kennes18} as well as to proposed $p+ip$ triplet superconductivity in Sr$_2$RuO$_4$ \cite{mackenzie03}. 
Since control of the order parameter relies solely on symmetry principles instead of materials specificities, details of the interaction and dissipation mechanisms as well as strong-coupling effects and retardation will not qualitatively change the results. Without loss of generality, we therefore numerically solve self-consistent Kadanoff-Baym equations with dissipation for low-energy quasiparticle models with effective attractive nearest-neighbor interactions that stabilize $d_{x^2-y^2}\pm id_{xy}$  and $(p_x \pm ip_y) \hat{\mathbf{z}}$ superconductivity in graphene and Sr$_2$RuO$_4$, respectively [see Methods]. The latter is theoretically expected to compete with helical triplet pairing states $p_x \hat{\mathbf{x}}$, $p_y \hat{\mathbf{y}}$ with trivial bulk topology \cite{scaffidi13}. We emphasize that our proposal permits to distinguish between the two in a pump-probe experiment.

\begin{figure*}[t]
	\centering
	\includegraphics[width=0.95\textwidth]{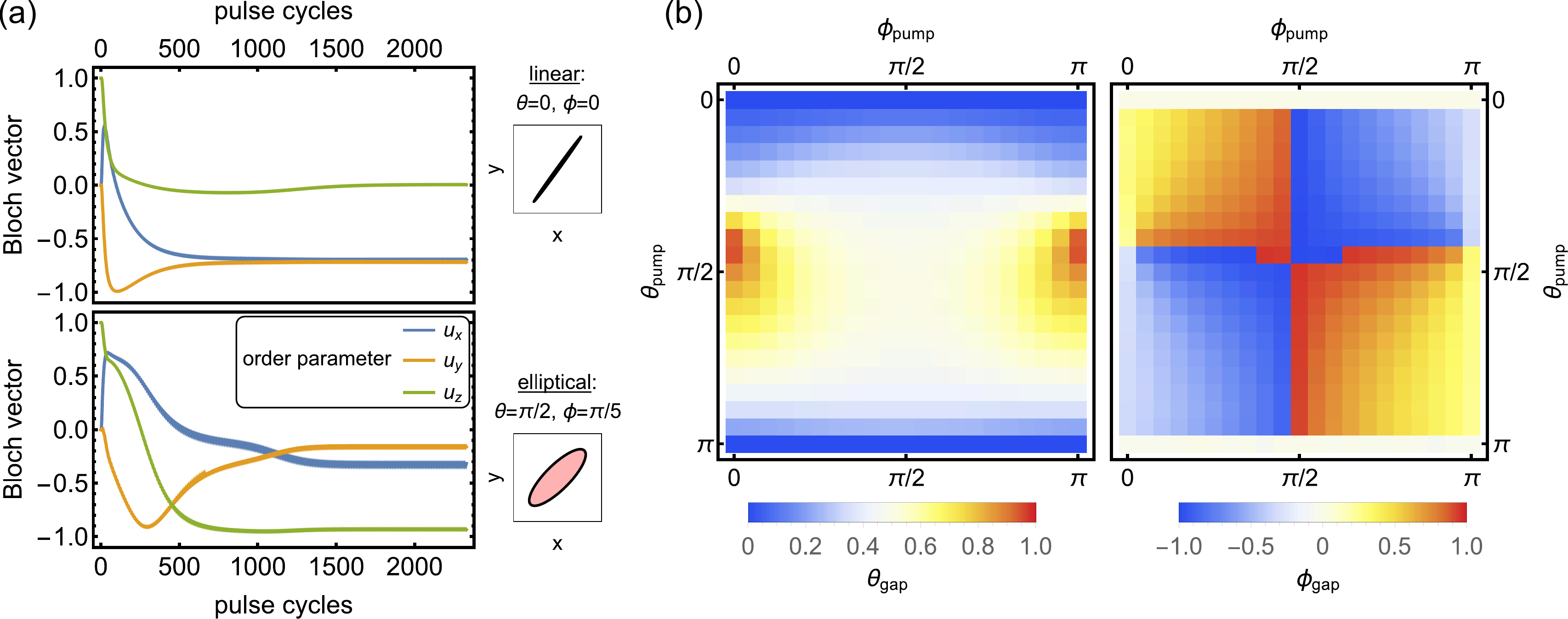}
	\caption{\captiontitle{Universal control of chiral superconductors.} (a) Time evolution and steady states for a high-frequency wide envelope pump pulse, starting from a $d+id$ state in graphene, for linear and elliptical polarization. The polarization ellipses are depicted as insets. (b) Dependence of the steady-state order parameter Bloch vector $\boldsymbol{\Delta} = \left[ \cos(\theta_{\rm gap}/2), ~\sin(\theta_{\rm gap}/2) e^{i\phi_{\rm gap}} \right]^\top$ on the choice of elliptical pump polarization, expressed via a Jones vector $[ \cos(\theta_{\rm pump}/2), ~\sin(\theta_{\rm pump}/2) e^{i\phi_{\rm pump}} ]^\top$. Linear-horizontal and circular polarization correspond to $\theta = 0$ and $\theta = \pi/2, \phi = \pi/2$, respectively. Each data point is computed as the time-evolved steady state, starting from a $d+id$ state. A simple physical picture emerges via noting that a generic elliptically-polarized pulse can always be represented as a superposition of pulses with linear and circular polarization; however, whereas rotational symmetry breaking due to linear polarization is proportional to the square of the field strength $\sim A^2$, the magnitude of mirror and time-reversal symmetry breaking is proportional to $\sim A^2 / \Omega$. The relative strength of symmetry-breaking perturbations subsequently determines the order parameter steady state. 
	}
	\label{fig:UniversalControl}
\end{figure*}

We first study the order parameter dynamics for a terahertz two-pulse sequence of linearly and circularly polarized pulses, red-detuned from the superconducting gap to suppress photo-induced pair breaking. Fig. \ref{fig:UltraFast} (a)-(c) depicts the time evolution of the superconducting gap magnitude and Bloch vector, starting from a $d_{x^2-y^2} + id_{xy}$ state in graphene, as a function of pump strength. The equilibrium superconducting gap in these simulations is amplified for numerical ease.

As expected from symmetry arguments, the first pulse with linear polarization rotates the Bloch vector from $\blochVec{0}{0}{1}$ ($d_{x^2-y^2} + id_{xy}$) towards nodal $d_{x^2-y^2}$ pairing $\blochVec{1}{0}{0}$. As sub-gap pumping necessarily heats the superconductor via residual multi-photon processes, the gap magnitude is transiently suppressed, but recovers once the system returns to thermal equilibrium via dissipation. Coupling to the extended $s$-wave pairing channel remains negligible in all parameter regimes studied here. The dynamics can thus be thought of as effective non-linear dissipative Bloch equations for the order parameter Bloch vector. Crucially, the second circularly-polarized pulse triggers by choice of handedness a change in chirality beyond a critical pump strength, rotating the order parameter towards $\blochVec{0}{0}{-1}$ ($d_{x^2-y^2} - id_{xy}$) [Fig. \ref{fig:UltraFast}(c)]. 

Fig. \ref{fig:UltraFast}(d)-(f) depicts the same pulse sequence for a three-orbital effective model of the $t_{\textrm{2g}}$ orbitals of Sr$_2$RuO$_4$. Analogously, the triplet-paired state rotates from $p+ip$ to $p-ip$ beyond a critical pump strength. We stress that this qualitative universality of the order parameter dynamics highlights the nature of the photo-induced control of chirality as a consequence of selective symmetry breaking alone, making it applicable to any chiral condensate with spontaneous TRS breaking. Finally, Fig. \ref{fig:UltraFast}(g) depicts the switching of the Bloch vector as in Fig. \ref{fig:UltraFast}(c) but represented on the Bloch sphere [see Supplementary Information for a movie of the induced dynamics].

\begin{figure*}[t]
	\centering
	\includegraphics[width=\textwidth]{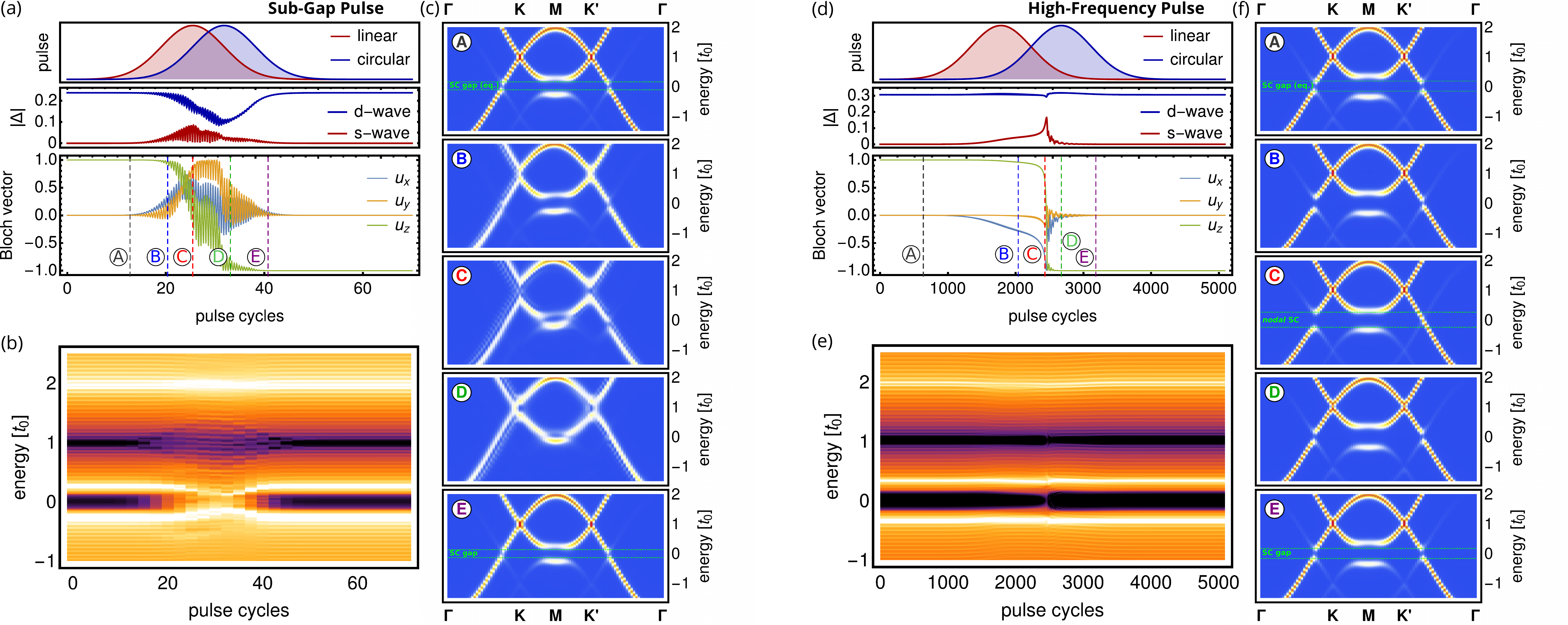}
	\caption{\captiontitle{Pump-probe measurement of order parameter dynamics.} Order parameter evolution [(a), (d)], time-dependent local density of states [(b), (e)],  and time-resolved spectral functions [(c), (f)], for a two-pulse sequences that reverses the chirality of a $d+id$ state in graphene. Left [(a)-(c)] and right [(d)-(f)] panels depict sub-gap ($\omega = 0.67 \Delta_{\textrm{eq}}$) and high-frequency pumping ($\omega = 8.0 t_{\textrm{hop}}$), respectively. Transient spectral functions (c), (f) are depicted for time slices indicated in (a), (d), for a probe pulse width of $\sigma_{\textrm{probe}}=10$ [see main text]. Density of states and spectral functions [see Methods] depict symmetry breaking and closing of the single-particle gap. For high-frequency pumping, a nodal state becomes cleanly visible along the $\Gamma \to K'$ cut at the topological transition between $d+id$ and $d-id$, with a V-shaped local density of states. In general, transiently-induced nodes of the superconducting gap are not constrained to lie on high-symmetry cuts, requiring a mapping of the entire Fermi surface.} 
	\label{fig:spectra}
\end{figure*}

To distinguish the roles of dynamical symmetry breaking and heating, we compared different pulse widths [see Supplementary Information, Fig. S1]. As expected, energy absorption increases for a longer pulse, suppessing the superconducting gap. However, the Bloch vector dynamics remain qualitatively unchanged. Albeit applied to few cycle pulses, this observation is consistent with expectations from Floquet theory, where dynamical crystal and TRS breaking enters as order $\sim A^2$ \cite{dunhap86} and $\sim A^2 t_{\textrm{hop}} / \Omega$ \cite{oka09} perturbations to the Hamiltonian, respectively. However, notably these remain much smaller than the equilibrium superconducting gap for the weak pulses considered, in contrast to the regime of high-frequency pumping [see Supplementary Information, Fig. S2]. Transient energy absorption is thus crucial in permitting to rotate the superconducting chirality already for weak critical pulse strengths $A$, and can be effectively understood as transitioning through an intermediate state far from thermal equilibrium.

Having established a protocol to optically reverse the handedness of a chiral superconductor and associated edge modes, a natural follow-up question is whether such universal optical control of the order parameter Bloch vector can transiently induce an arbitrary order parameter. To this end, elliptical polarization provides a natural generalization to access the entire state space. Consider a $d+id$ state in graphene, irradiated by a wide high-frequency pulse with polarization Jones vector $\left[ \cos\left( \frac{\theta}{2} \right),~ \sin\left( \frac{\theta}{2} \right) e^{i\phi} \right]^\top$. Fig.~\ref{fig:UniversalControl} depicts the induced order parameter dynamics and long-time steady state as a function of polarization vector. Away from linear ($\theta = 0$) or circular ($\theta,\phi = \pi/2$) polarization, the laser pulse now jointly breaks lattice rotation, in-plane mirror and time-reversal symmetries. The transient order parameter eigenstates cease to be $d_{x^2-y^2}$, $d_{xy}$ or $d_{x^2-y^2} \pm i d_{xy}$, instead forming a linear superposition that reflects the polarization of the pump. The two angular degrees of freedom of pump pulse polarization thus suffice to permit universal control of the vectorial order parameter.

Beyond direct observation of the induced Majorana edge modes along domain wall boundaries in transport or via spectroscopic methods, one possible experimental confirmation of our predictions follows from noting that the switching from $d+id$ to $d-id$ ($p+ip$ to $p-ip$) entails a bulk topological transition, necessitating the transient closing of the superconducting gap at discrete points on the Fermi surface. The time-resolved single-particle spectral function and local density of states are particularly simple proxies, measurable in pump-probe angle-resolved photoemission or time-resolved STM with picosecond time resolution for typical switching pulse durations.

Fig. \ref{fig:spectra} depicts the simulated pump-probe response for $d+id$ pairing in graphene [see Methods]. For terahertz pumping, the local density of states [Fig. \ref{fig:spectra}(b)] exhibits a closing of the single-particle spectral gap near the topological transition between $d+id$ and $d-id$. The corresponding spectral function [Fig. \ref{fig:spectra}(c)] shows the transient breaking of rotation symmetry, with a unidirectional gap closure in the middle panel at the topological transition. The occurrence of a nodal state at the topological transition is partially masked by frequency-shifted side bands as well as transient heating, which uniformly suppresses the superconducting gap.

Conversely, in the high-frequency regime side bands are shifted to high energies and a clean transition from a chiral $d+id$ to a nodal $d$-wave superconductor and back to a $d-id$ state with opposite chirality is observed [see Fig. \ref{fig:spectra}(d), (e), (f)]. A node appears on the Fermi surface along the $K' \leftrightarrow \Gamma$ cut. Analogously, the local density of states acquires a V-shaped form at the topological transition as expected for a nodal state, before returning to a fully-gapped chiral state. In addition, the extended $s$-wave pairing symmetry acquires a finite magnitude. However, a transient closing of the single-particle gap remains unavoidable for realizing a topological transition, hence serving as direct evidence of the topological character of the superconducting state.

An immediate consequence of our work is the possibility to optically engineer and control arbitrary chiral domains on ultrafast time scales, paving the way to use local optical control of the order parameter to selectively induce or destroy domain-wall Majorana modes. Intriguingly, while the local order parameter itself cannot encode a qubit, such optically-induced domain-wall Majorana fermions could be used to perform topological quantum computation via electrical injection and read-out, in analogy to superconductor - quantum anomalous Hall insulator junctions \cite{lian2018}. Physical qubits for computation are represented by charging states of single-electron-injected wave packets, while the sequence of braiding operations is controlled via the topography of chiral domains, which in turn can be defined optically using ultrafast pulses. Such a combined experiment of optically-controlled topological quantum gates would constitute a key advance toward a programmable topological quantum computer.

Second, as the presented mechanism relies solely on a two-dimensional irrep of the order parameter, analogous predictions apply to particle-hole condensates such as $d+id$ density waves in hexagonal systems \cite{nayak00}. There, unlike previously-proposed Floquet engineering of topological materials \cite{lindner2011floquet}, our proposal permits ultrafast optical switching of the quantized Hall conductance and charge-carrying dissipationless chiral edge modes, which persist after the pump is switched off.

\hspace{0.5in}

\bibliographystyle{naturemag}


\appendix*

\section*{Methods} 

We employ an effective honeycomb lattice model of the lowest-energy quasi-flat bands in TBG at $\sim$meV kinetic energy scales \cite{mcdonaldPNAS2011,yuan18}, as well as an effective three-orbital quasiparticle model for the t$_{\textrm{2g}}$ orbitals in Sr$_2$RuO$_4$ with spin-orbit coupling. Additionally, optical pumping necessarily heats the superconductor, hence the electronic system must dissipate heat to the substrate and environment in order to return to equilibrium at long times. Since we are primarily interested in the thermalized state \textit{after} the pump instead of details of the relaxation dynamics, the specifics of dissipation are irrelevant. A computationally advantageous choice is to consider tunneling to a wide-band electronic substrate, while ignoring dissipation via electron-phonon coupling. With these considerations, the self-consistent Kadanoff-Baym equations for irradiated chiral superconductors are solved numerically.

\paragraph*{Details of Graphene Calculation}

A decade after first proposals that electronic interactions in heavily-doped monolayer graphene can induce effective attractive pairing for chiral $d$-wave superconductivity at low energies \cite{blackschaffer06,nandkishore11,kiesel11}, experiments found evidence of superconductivity in gated TBG at magic twist angles \cite{yuan18}. Among a host of proposed theories, the common theme is that if the superconducting state indeed derives from repulsive electronic interactions in a flat band, then the dominant instability should be chiral pairing in the $d$-wave channel \cite{xu18,dodaro18,guo18,huang18,liu18,kennes18,po18,isobe18}, although other mechanisms such as standard electron-phonon coupling have been proposed as well \cite{padhi18,lian18}.

Following Refs.~\cite{mcdonaldPNAS2011,yuan18}, the lowest-energy quasi-flat bands in TBG can be captured via an effective two-orbital honeycomb lattice model at $\sim$meV kinetic energy scales.
\begin{align}
	\Ham = -\sum_{\substack{nn' \\ \alpha\alpha'\sigma }} t_{nn'}^{\alpha\alpha'}~  \CD{n\alpha\sigma} \C{n'\alpha'\sigma} - \mu \sum_{i\alpha\sigma} \ND{i\alpha\sigma}
\end{align}
	where $\alpha,\alpha'$ and $\sigma$ denote orbital and spin indices.
	Upon neglecting weak breaking of SU(4) sublattice symmetry \cite{kennes18}, $t_{nn'}^{\alpha\alpha'} \neq 0$ only for nearest-neighbor hopping between the same orbitals, and mean field theory for effective renormalized interactions becomes identical to monolayer graphene doped to the van-Hove singularity \cite{nandkishore11,kiesel11}, but with an additional orbital degree of freedom.

Since control of the order parameter relies solely on symmetry principles instead of materials specificities, microscopic details of the interaction as well as strong-coupling effects and retardation should not qualitatively change the results. Without loss of generality, we therefore constrain our simulations to nearest-neighbor attractive interactions
$
V_\q = -V \gamma_\q \hat{\tau}^+ + \hc 
$ that are orbitally isotropic, with 
\begin{equation}
\gamma_\k = 1 + e^{-i ( k_x/2 - \sqrt{3} k_y/2 )} + e^{+i ( k_x/2 + \sqrt{3} k_y/2 )}
\end{equation} and $\hat{\tau}^+ \equiv \hat{\tau}^x + i \hat{\tau}^y$ the sublattice Pauli matrices. Such interactions stabilize $d_{x^2-y^2}$ and $d_{xy}$ pairing with 
\begin{equation}\hat{\eta}_\k^{(1)} = \frac{1}{\sqrt{6}} \left[ 2 - e^{i(k_x/2 - \sqrt{3} k_y/2)} - e^{-i(k_x/2 - \sqrt{3}k_y/2)} \right] \hat{\tau}^+ + \hc 
\end{equation}
 and 
 \begin{equation} 
 \hat{\eta}_\k^{(2)} = \frac{1}{\sqrt{2}} \left[  e^{i(k_x/2 - \sqrt{3} k_y/2)} - e^{-i(k_x/2 - \sqrt{3}k_y/2)} \right] \hat{\tau}^+ + \hc, 
 \end{equation}
that span the basis for the order parameter Bloch vector. Furthermore, extended $s$-wave pairing with 
 \begin{equation}\hat{\eta}_\k^{(s)} = \frac{1}{\sqrt{3}} \left[ 1 + e^{i(k_x/2 - \sqrt{3} k_y/2)} + e^{-i(k_x/2 - \sqrt{3}k_y/2)} \right] \hat{\tau}^+ + \hc  \end{equation}
 is permitted by symmetry but suppressed in equilibrium. This choice of $V_\q$ matches expectations for renormalized interactions from functional renormalization group calculations \cite{kennes18}.
 
 The pump field enters via Peierls substitution, with hopping amplitudes $t_{nn'} \to t_{nn'} e^{i \int_{n'}^{n} d\mathbf{r} \cdot \mathbf{A}(t)}$. The dimensionless peak pulse strengths quoted in the main text relate to the electric field $\mathcal{E}$ via $A = a_0 e \mathcal{E} / (\hbar \Omega)$, where $a_0$, $e$ are the lattice constant and electric charge, respectively.
 The order parameter simulations for graphene in the main text were performed for $V=-0.621$, $\mu=-0.95$, $\Gamma=0.05$ at bath temperature $T=0.1$, yielding an equilibrium gap of $\Delta \sim 0.1$, all in units of nearest-neighbor hopping $t_{\textrm{hop}}$. Here, $\Gamma$ denotes the relaxation rate to a wide-band electronic substrate [see Supplementary Information, section IV], Temperature, relaxation constant and superconducting gap are all exaggerated to ease numerical simulations. High-frequency simulations were performed for pump frequencies of 1.3 times the electronic bandwidth. Simulations of the pump-probe response were performed for $T=0.05$, $\Gamma=0.01$ and $V=-0.067$, at signficantly increased numerical cost.

\paragraph*{Details of Sr$_2$RuO$_4$ Calculation} 

To emphasize the universality of the order parameter dynamics, we additionally study the evolution of the chiral gap function in a low-energy quasi-particle model of Sr$_2$RuO$_4$. We employ a three orbital tight-binding model of maximally-localized 
Wannier functions \cite{Souza_2001} constructed for the
Ru-4d t$_{\textrm{2g}}$ subspace from a non-spin-polarized density functional 
calculation (using Wien2k~\cite{Wien2k} with PBE functional~\cite{PBE},
wien2wannier~\cite{wien2wannier} and wannier90~\cite{wannier90}). The 
local self-energy is obtained in the normal state (at 29K)
with dynamical mean-field theory~\cite{Georges_1996} using the TRIQS 
toolbox~\cite{TRIQS,TRIQS/CTHYB,TRIQS/DFTTools}.
Additionally, we add a single particle spin-orbit term with an enhanced 
coupling constant of 200meV. This choice is based on theoretical
studies~\cite{Zhang_2016, Kim_2018} and justified by an in-depth 
analysis of high-precision angle-resolved photo-emission
measurements~\cite{Tamai_tbp}. In the latter, details of the model and dynamical mean-field theory calculation are presented. A two-dimensional quasi-particle model is derived via linearization of the self-energy at the Fermi surface. This effective model describes quasi-particles in a $\sim \pm 30\textrm{meV}$ energy range around the Fermi energy, hence underestimating the overall electronic bandwidth. However, we expect that, for our purposes, this is not pertinent to the two off-resonant regimes under consideration. For sub-gap pumping, the time evolution can be expected to be sensitive only to states in the vicinity of the Fermi energy, whereas the high-frequency Floquet regime is essentially adiabatic, with the bare bandwidth $W$ disappearing as a relevant energy scale provided that the frequency $\Omega \gg W$.

With these considerations in mind, we truncate the effective quasiparticle tight-binding model beyond 10$^{\textrm{th}}$-nearest neighbor hopping:
\begin{align}
	\mathbf{H}_\k &= \left[\begin{array}{ll} \mathbf{h}_{\k,+} & 0 \\ 0 & \mathbf{h}_{\k,-} \end{array}\right], \notag\\
	\mathbf{h}_{\k,\nu} &= \left[\begin{array}{ccc}
			\E_\k^{\textrm{xz}} & -i \nu \lambda + g_\k & i \lambda \\
			i \nu \lambda + g_\k & \E_\k^{\textrm{yz}} & -\nu \lambda \\
			-i \lambda & -\nu \lambda & \E_\k^{\textrm{xy}}
		\end{array}\right]
\label{eq:Sr2RuO4model}
\end{align}
with $\nu = \pm$, where $\lambda$ is the renormalized spin-orbit coupling constant. The kinetic matrix elements and the corresponding Fermi surface depicting the orbital character of the individual sheets are provided in the Supplementary Information, section III.

The tight-binding model transforms under the point group D$_{\textrm{4h}}$, and hence has a degeneracy (two-fold irreducible representation E$_\textrm{u}$) between triplet $p_x \hat{\mathbf{z}}$, $p_y \hat{\mathbf{z}}$ states, where $\mathbf{z}$ refers to the Pauli matrix in the usual triplet channel spinor representation. It is invariant under inversions, as well as $C_4$ rotations. For the main text, we consider an attractive interaction in the $(p_x + ip_y) \hat{\mathbf{z}}$ pairing channel. We note that while the nature of the superconducting order is under active investigation \cite{Mackenzie2017}, chiral pairing can be distinguished from competing proposals in a pump-probe experiment as shown in the main text. Further symmetry considerations are deferred to the Supplementary Information.

\paragraph*{Details of the Order Parameter Evolution}

We start from the self-consistent Kadanoff-Baym equations for the Green's function $\mathcal{G}$ in Keldysh-Nambu basis 
\begin{align}
	i \partial_t \mathcal{G}_\k(t,t') &= \mathcal{H}_\k(t, \boldsymbol{\Delta}_\k(t) ) ~  \mathcal{G}_\k(t,t') \notag\\
		&+ \int d\tau~ \hat{\Sigma}_\k(t,\tau)~ \mathcal{G}_\k(\tau,t') \\
	\boldsymbol{\Delta}_\k(t) &= \frac{1}{L} \sum_{j} v^{(j)} \hat{\boldsymbol{\eta}}_{\k}^{(j)} \sum_{\substack{\k' \\ \alpha\beta}} \hat{\eta}_{\k'\alpha\beta}^{(j)} \expect{ \C{-\k',\beta\downarrow} \C{\k',\alpha\uparrow} }
\end{align}
and solve the resulting integro-differential equations of motion on a $60\times 60$ momentum point grid with a time step no larger than $1/40$ of the pump period.
Here, $\mathcal{G}_\k(t,t')$ is the Keldysh-Nambu Green's function, $\mathcal{H}_\k(t, \boldsymbol{\Delta}_\k(t) )$ denotes a multi-band Bogoliubov de Gennes Hamiltonian, and the pump field couples to electrons via Peierls substitution. Attractive interactions are treated in an instantaneous mean field approximation for the superconducting multi-band order parameter $\boldsymbol{\Delta}_\k$, with orbital indices denoted by $\alpha,\beta$, and where $v^{(j)}$ denotes the decomposition of two-body interaction matrix elements in terms of form factors $\hat{\eta}_{\k}^{(j)}$ for the relevant irreducible representations, indexed by $j$. Finally, the self energy $\hat{\Sigma}$ models weak dissipative coupling to a metallic substrate. In the wide-band limit for the bath, this corresponds to a retarded self energy $\Sigma^R(t,t') = -i \Gamma~ \delta(t-t')$ and Keldysh self-energy $\Sigma^K(t,t') = -2\Gamma ~T \left[ \sinh(\pi (t-t') T) \right]^{-1}$, with $\Gamma$ and $T$ the effective relaxation constant and substrate equilibrium temperature, respectively.
Further details can be found in the Supplementary Information, section IV. 

\paragraph*{Details of the $\rho(\omega,T)$ and $A_\k(\omega,T)$ Calculation}
 
We calculate the time-resolved local density of states

\begin{equation}
\rho(\omega,T)=2 \textrm{Im} \int d\tau~ e^{-i\omega \tau} \tr \mathbf{G}^R(T+\tau/2, T-\tau/2)
\end{equation} as well as the probe-averaged single-particle spectral function \begin{equation}
A_\k(\omega,T)
=\int dt_1 dt_2~ e^{-i\omega(t_1-t_2)} s(t_1) s(t_2) \tr \mathbf{G}^R(t_1,t_2)
\end{equation}
using a Gaussian probe pulse shape function $s(\tau) = e^{-(\tau-T)^2/(2 \sigma_{\textrm{probe}})}$ and where $ \mathbf{G}^R$ denotes the retarded Green's function. We employ a momentum grid with $90 \times 90$ points, and the equilibrium superconducting gap is exaggerated to ease the computational cost without changing the qualitative behavior.

\acknowledgements{
We gratefully thank Antoine Georges and Andrew J. Millis for helpful discussions. M.C. and M.Z. are supported by the Flatiron Institute, a division of the Simons Foundation. D.M.K. and M.A.S. acknowledge support from the DFG through the Emmy Noether program (KA 3360/2-1 and SE 2558/2-1, respectively). We acknowledge financial support from the European Union Horizon 2020 research and innovation program under the European Research Council (ERC Advanced Grant Agreement no. 69409). 

\paragraph{Contributions:} M.C. conceived the idea and performed the time-domain calculations. M.C., D.M.K. and M.A.S. analyzed the results. M.Z. performed the DFT+DMFT simulations. A.R. supervised the project. All authors contributed to discussions and to the writing of the manuscript.

\paragraph{Competing financial interests:}
The authors declare no competing financial interests.

\paragraph{Corresponding author:}
Martin Claassen \\ (mclaassen@flatironinstitute.org).

\paragraph{Data availability:}
All data generated and analyzed during this study are available from the corresponding author upon request.
}

\clearpage
\pagebreak

\section*{Supplementary Information}
\renewcommand\thefigure{S\arabic{figure}}
\setcounter{figure}{0}
\renewcommand\theequation{S\arabic{equation}}
\setcounter{equation}{0}

\subsection*{Pump Envelope Dependence}

\begin{figure*}[t]
	\centering
	\includegraphics[width=\textwidth,trim=100pt 150pt 60pt 200pt]{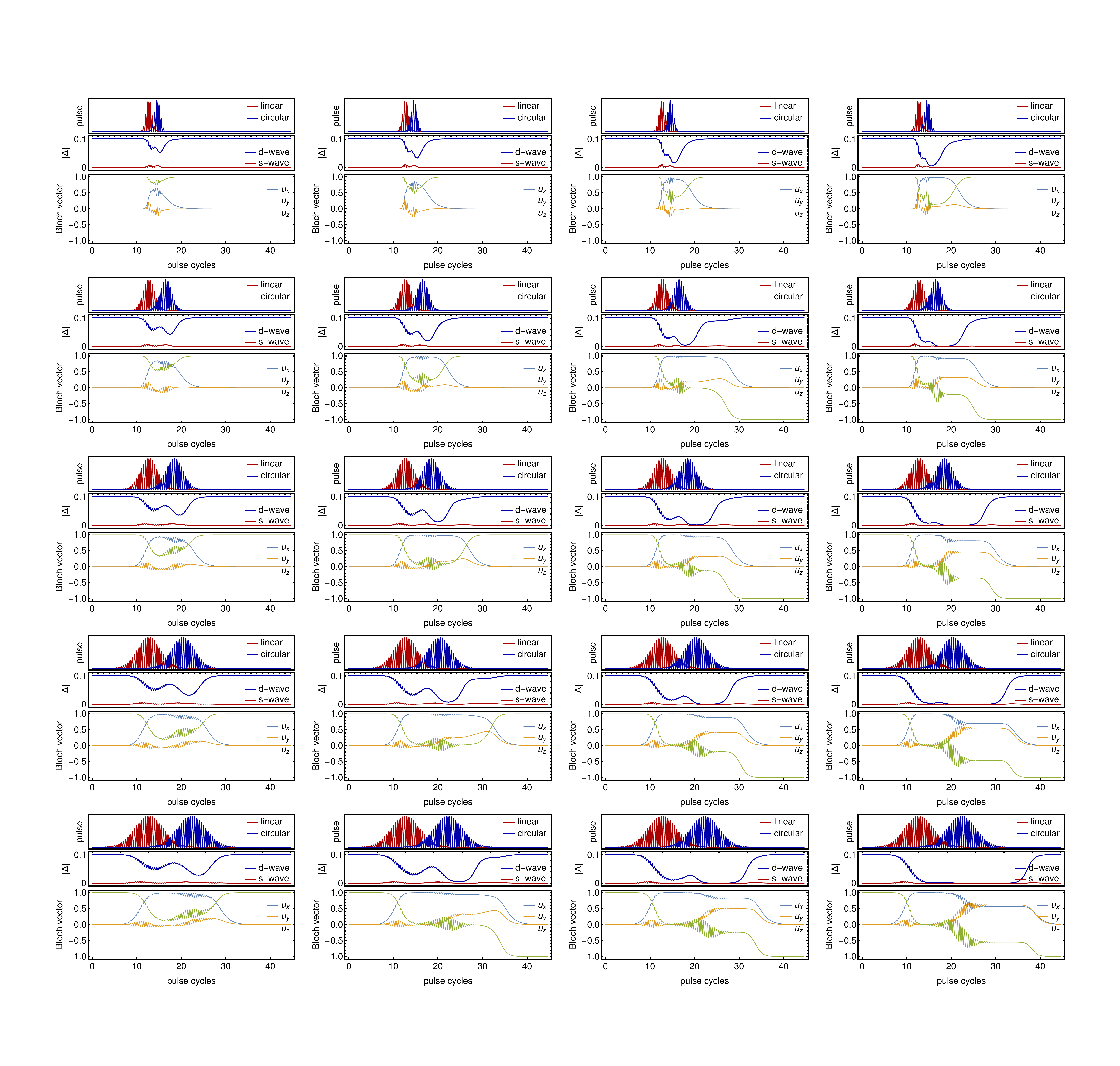}
	\caption{\captiontitle{Pulse width dependence of order parameter dynamics.} Order parameter dynamics for a $d+id$ superconductor in graphene as a function of increasing pulse width with $\Omega = 0.8|\Delta|(t=0)$, from top to bottom [depicted in the top sub-panels], and pump strength $A=0.25,0.3,0.35,0.4$, from left to right. Increasing the pulse width increases energy absorption and subsequently quenches superconducting order. Conversely, symmetry breaking and Bloch vector dynamics are largely independent of pulse width.} 
	\label{fig:EnvelopeDependence}
\end{figure*}

Here, we briefly discuss the dependence of the chiral order parameter dynamics on the width of the pump pulse. As discussed in the main text, the ensuing dynamical symmetry-breaking is best understood in the Floquet limit of wide envelope pulses, however optical reversal of chirality similarly happens for ultrafast optical pulses.

It is therefore desirable to demonstrate that the dynamics in both limits are qualitatively equivalent. Fig. \ref{fig:EnvelopeDependence} depicts the dynamics of the gap magnitude and Bloch vector for a driven $d+id$ superconductor in graphene, as a function of pulse width. One finds that the superconducting gap magnitude is quenched for increasing pulse widths [top to bottom rows], however the dynamics of the Bloch vector are largely unaffected, instead depending solely on the peak field strength [left to right columns].

The dependence of the Bloch vector dynamics on peak field strength, as well as the lack of dependence on pulse width beyond a minimum duration, corresponds to expectations for the Floquet limit of wide pump pulses, carried over to ultrafast time scales. In the Floquet limit, the pulse can be viewed as a dynamical symmetry-breaking perturbation to the kinetic Hamiltonian that induces dynamics within the two-dimensional order parameter subspace as discussed in the main text. Dynamical reduction of the discrete lattice rotation symmetry generically enters as a perturbation proportional to $\sim \left[ \mathcal{J}_0(A) - 1 \right]$  the zeroth Bessel function, quadratic in the peak field strength at weak fields \cite{dunhap86,eckhard17}. Similarly, dynamical breaking of time-reversal symmetry appears as a perturbation $\sim \left[ \mathcal{J}_1(A) \right]^2 / \Omega$, again quadratic in the peak field strength while simultaneously sensitive to the pump frequency \cite{oka09,kitagawa11}. Conversely, the reduction of the superconducting gap for wider pulse envelopes can be understood as a result of energy absorption. For wider pulse envelopes, dissipation via coupling to the substrate ultimately stabilizes a Floquet steady state with suppressed superconducting gap.

\subsection*{Chirality Reversal in the High-Frequency Regime}

\begin{figure*}[t]
	\centering
	\includegraphics[width=0.9\textwidth]{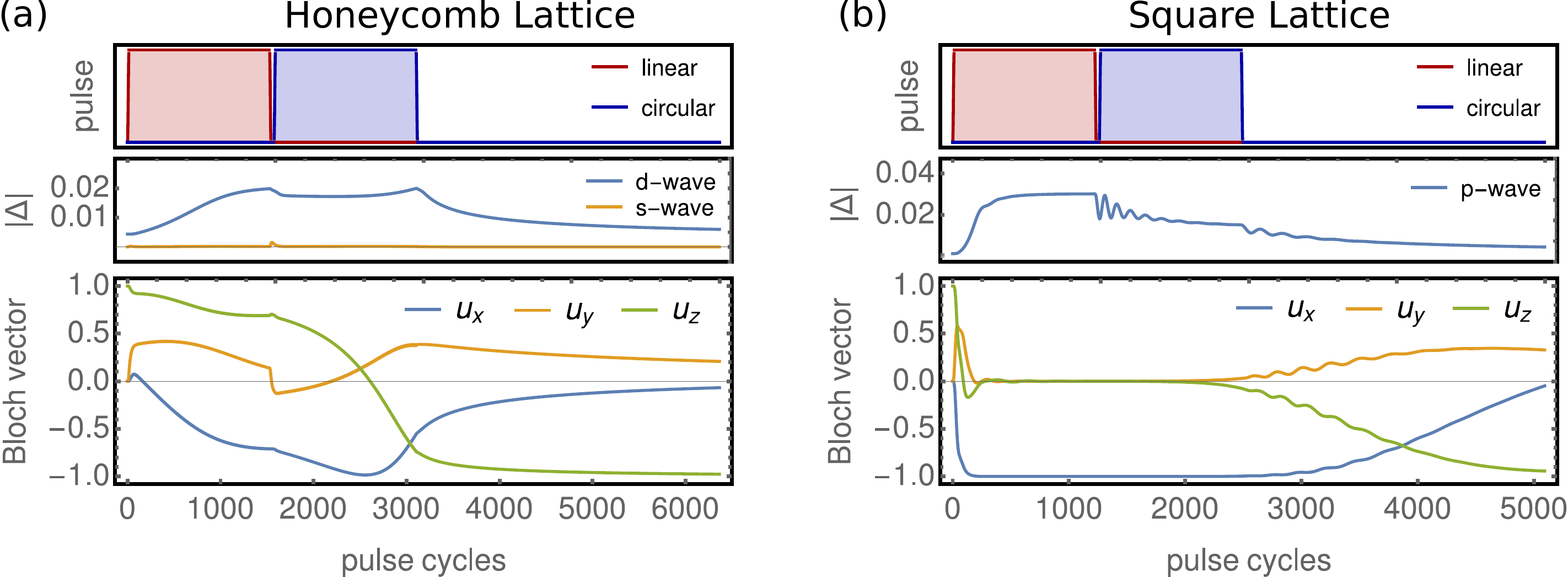}
	\caption{\captiontitle{Floquet reversal of the chirality of the order parameter in the high-frequency regime.} Proof-of-principles depiction of a controlled rotation of the superconducting chirality using a two-pulse sequence of wide high-frequency pulses in the Floquet regime. (a) and (b) depict the rotation of the order parameter Bloch vector for $p_x + i p_y$ and $d_{x^2-y^2} + i d_{xy}$ order on the square lattice and honeycomb lattice, respectively. Top, middle and bottom panels depict the pulse sequence, evolution of the superconducting gap magnitude and evolution of the order parameter Bloch vector, respectively. In analogy to the ultrafast subgap pulse sequences discussed in the main text, linear polarization drives the superconductor toward a nodal steady state; once the order parameter is liberated from the equilibrium free energy minimum, a second circularly-polarized pulse subsequently biases the evolution towards a chiral state with opposite chirality. Furthermore, in the high-frequency Floquet limit heating is suppressed, leading to transient enhancement of the magnitude of the superconducting gap due to dynamical flattening of the electronic bands \cite{dunhap86,dasari18}.}
	\label{fig:FloquetControl}
\end{figure*}

Analogous to the discussion of sub-gap pumping in the main text, a sequence of linearly-polarized and circularly-polarized high-frequency pulses can in principle lead to a controlled reversal of the chirality of the superconducting order. Fig. \ref{fig:FloquetControl}(a) depicts a $d+id$ singlet superconducting state on the honeycomb lattice, doped close to the van Hove singularities with $\mu = -0.95 t_{\rm hop}$, where $t_{\rm hop}$ denotes nearest-neighbor hopping, and driven with a wide-envelope two-pulse sequence of a linearly- and circularly-polarized pulse.  At high frequency $\Omega = 10 t_{\rm hop}$, a strong linearly-polarized pulse with $A=1$ transiently quenches hoppings along the $x$ direction proportional to the zeroth Bessel function of pump strength $\mathcal{J}_0(A)$ \cite{dunhap86}, unidirectionally increasing the density of states at the Fermi level and breaking the lattice rotation symmetry. The $d+id$ state ceases to represent a steady state of the system, and the pulse rotates from $d+id$ to $d_{x^2 - y^2}$.  A second circularly-polarized pulse subsequently rotates toward a chiral final state with opposite chirality $d-id$. Coupling to an extended $s$-wave state state is negligible for sufficiently smooth pump envelopes, and $s$-wave pairing does not become an accessible steady state solution over all investigated parameter regimes.

Unlike sub-gap pumping discussed in the main text, heating is suppressed on exponentially-long time scales for high-frequency pumping, and the rotation of the order parameter can be effectively understood in terms a transient \textit{static} pre-thermal Floquet Hamiltonian which can be derived from a Magnus expansion \cite{abanin15a,abanin15b}. Thus, the superconducting order parameter can be thought of effectively following the free energy minimum of the Floquet Hamiltonian, requiring much larger pulse strengths to induce a rotation of the chiral order parameter.

Similarly, Fig. \ref{fig:FloquetControl}(b) depicts the same two-pulse sequence for a $p+ip$ triplet superconducting state on the single-orbital square lattice, doped $\mu = -1/2 t_{\rm hop}$ away from half filling. At high pump frequencies, a strong linearly-polarized pulse analogously transiently quenches superconductor to a nodal $p_x$ paired state, a second circularly-polarized pulse subsequently picks the chirality of the final state. Here, dynamical breaking of time-reversal symmetry is a higher-order effect in $A$ -- an artifact of a single-band model -- nevertheless, rotation of the chirality of the order parameter proceeds analogously.


\subsection*{Low-Energy Model for the $d_{xz}$, $d_{yz}$, $d_{xy}$ orbitals of $\textrm{Sr}_2\textrm{RuO}_4$}

\begin{figure}[t]
	\centering
	\includegraphics[width=0.9\columnwidth, trim= 15pt 0pt 0pt 0pt]{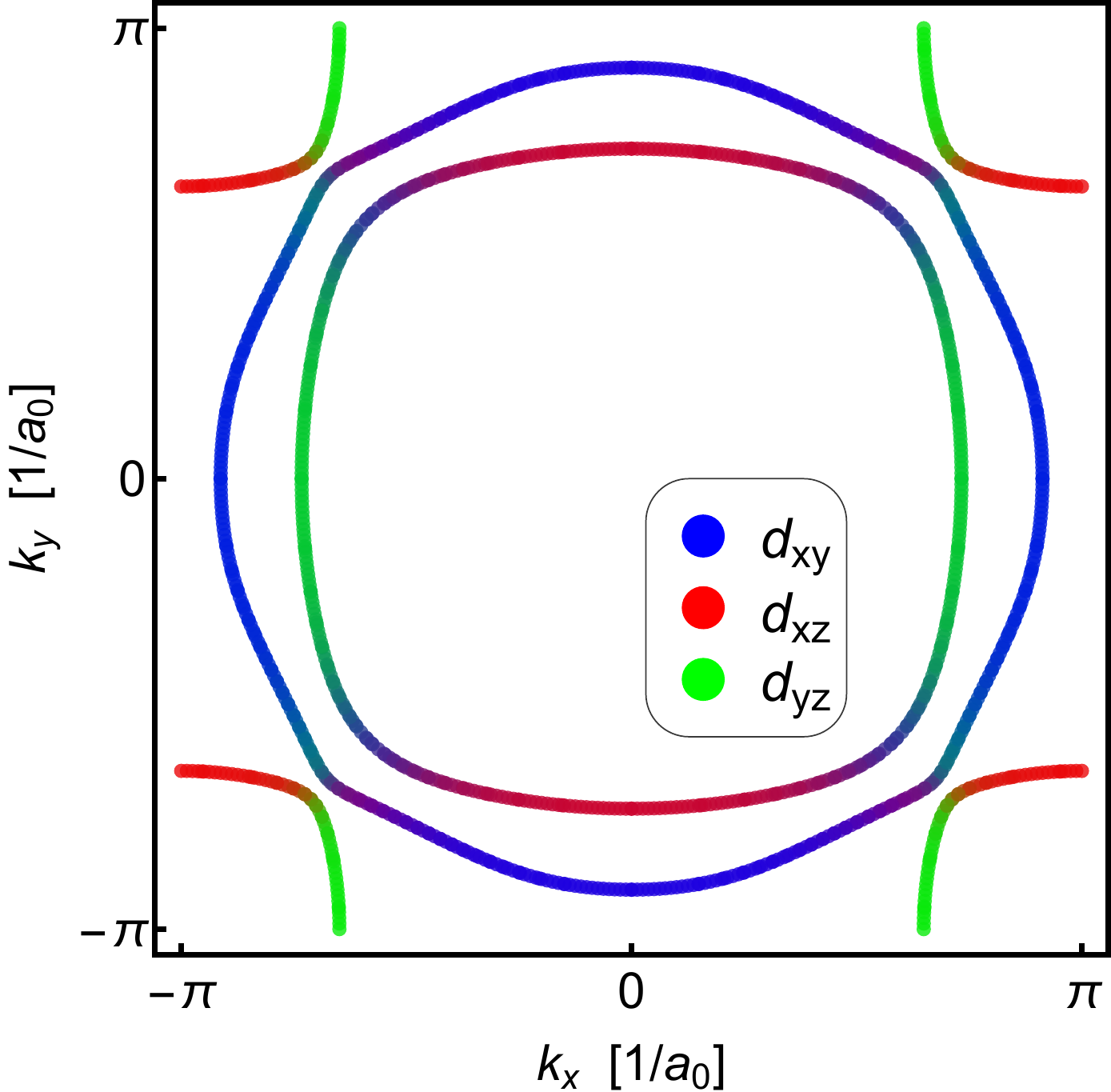}
	\caption{\captiontitle{Fermi surface and orbital character of Sr$_2$RuO$_4$.} Fermi surface and $d_{xy}$, $d_{xz}$, $d_{yz}$ orbital components, calculated from the effective quasi-particle tight-binding model of Eq. (\ref{eq:Sr2RuO4modelSupp}). The inclusion of spin-orbit coupling has a strong influence on the topography of the Fermi surface and also leads to a mixing of orbital characters along the individual sheets. The Fermi surface of this model is in excellent agreement with high-precision angle-resolved photo-emission data~\cite{Tamai_tbp}. }
\end{figure}

As discussed in the Methods section of the main text, we employ an effective two-dimensional three-orbital model of Sr$_2$RuO$_4$. Taking into account solely atomic spin-orbit coupling, symmetry considerations constrain the quasi-particle Hamiltonian to
\begin{align}
	\mathbf{H}_\k = \left[\begin{array}{ll} \mathbf{h}_{\k,+} & 0 \\ 0 & \mathbf{h}_{\k,-} \end{array}\right]
	\label{eq:Sr2RuO4modelSupp}
\end{align}
with
\begin{align}
\mathbf{h}_{\k,\nu} = \left[\begin{array}{ccc}
			\E_\k^{\textrm{xz}} & -i \nu \lambda + g_\k & i \lambda \\
			i \nu \lambda + g_\k & \E_\k^{\textrm{yz}} & -\nu \lambda \\
			-i \lambda & -\nu \lambda & \E_\k^{\textrm{xy}}
		\end{array}\right]
\end{align}

\begin{widetext}
Here we give the kinetic matrix elements used in the main text:
\begin{align}
	\E_\k^{\textrm{xz}} &= -0.109297 - 0.034634 \cos \left(k_x\right) - 0.195441 \cos \left(k_y\right)  + 0.023565 \cos \left(2 k_y\right)  \notag\\
	&+ 0.001559 \left[ \cos \left(k_x-3 k_y\right) +  \cos \left(k_x+3 k_y\right) \right] + 0.006043 \left[ \cos \left(k_x-2 k_y\right) + \cos \left(k_x+2 k_y\right) \right] \notag\\
	&+ 0.002895 \left[ \cos \left(k_x-k_y\right) + \cos \left(k_x+k_y\right) \right] + 0.001013 \left[ \cos \left(2 k_x-k_y\right)  + \cos \left(2 k_x+k_y\right) \right]  \notag\\
	& + 0.002835 \cos \left(2 k_x\right) + 0.002595 \cos \left(3 k_y\right) \\
	\E_\k^{\textrm{yz}} &= \E_{\left(k_y,-k_x\right)}^{\textrm{xz}} \\
	\E_\k^{\textrm{xy}} &= -0.074803 - 0.134663 \left[ \cos \left(k_y\right) + \cos \left(k_x\right) \right] + 0.001638 \left[ \cos \left(2 k_y\right) + \cos \left(2 k_x\right) \right] \notag\\
		&-0.044290 \left[ \cos \left(k_x-k_y\right) + \cos \left(k_x+k_y\right) \right] - 0.001678 \left[ \cos \left(3 k_x\right) + \cos \left(3 k_y\right) \right] \notag\\
		&- 0.007565 \left[ \cos \left(k_x-2 k_y\right) + \cos \left(2 k_x-k_y\right) + \cos \left(2 k_x+k_y\right) + \cos \left(k_x+2 k_y\right)  \right] \notag\\
		&-0.001545 \left[ \cos \left(k_x-3 k_y\right) + \cos \left(3 k_x+k_y\right) + \cos \left(3 k_x-k_y\right) + \cos \left(k_x+3 k_y\right) \right] \notag\\
		&- 0.001888 \left[ \cos \left(2 k_x-3 k_y\right) + \cos \left(3 k_x+2 k_y\right) \right] - 0.001375 \left[ \cos \left(3 k_x-3 k_y\right) + \cos \left(3 k_x+3 k_y\right) \right] \notag\\
		&-0.005025 \left[ \cos \left(2 k_x-2 k_y\right) + \cos \left(2 k_x+2 k_y\right) \right] - 0.001888 \left[ \cos \left(3 k_x-2 k_y\right) + \cos \left(2 k_x+3 k_y\right) \right] \\
	g_\k &= 0.007249 \left[ \cos \left(k_x-k_y\right) - \cos\left(k_x+k_y\right) \right] \notag\\
		&+0.001166 \left[ \cos \left(k_x-3 k_y\right) - \cos \left(k_x+3 k_y\right) + \cos \left(3 k_x-k_y\right) - \cos \left(3 k_x+k_y\right) \right] \\
	\lambda &= 0.028571
\end{align}

As mentioned in the main text the tight-binding model transforms under the point group D$_{\textrm{4h}}$. It is invariant under inversions, as well as $C_4$ rotations with
\begin{align}
	\left[\begin{array}{cccccc}  0 & 1 \\ -1 & 0  \\ & & i \\ & & & 0 & 1 \\ & & & -1 & 0 \\ & & & & & -i \end{array}\right]^{-1} \mathbf{H}_{(-k_y,k_x)} \left[\begin{array}{cccccc}  0 & 1 \\ -1 & 0  \\ & & i \\ & & & 0 & 1 \\ & & & -1 & 0 \\ & & & & & -i \end{array}\right] = \mathbf{H}_\k
\end{align}
and $\sigma_v$ reflections with
\begin{align}
	\left[\begin{array}{cccccc} 0 & & & 1 \\ & 0 & & & -1 \\ & & 0 & & & 1 \\ 1 & & & 0 \\ & -1 & & & 0 \\ & & 1 & & & 0\end{array}\right]^{-1} \mathbf{H}_{(-k_x,k_y)} \left[\begin{array}{cccccc} 0 & & & 1 \\ & 0 & & & -1 \\ & & 0 & & & 1 \\ 1 & & & 0 \\ & -1 & & & 0 \\ & & 1 & & & 0\end{array}\right] = \mathbf{H}_\k
\end{align}
and the corresponding rotated $\sigma_v$ and $\sigma_d$ mirror planes.
\end{widetext}

\subsection*{Self-Consistent Keldysh Formalism for the Order Parameter Dynamics}

\subsubsection*{Model}

In the main text, we consider a generic driven multi-band model with attractive interactions and coupled to a fermionic bath. To this end, a generic starting point reads
\begin{align}
	\Ham &= \Ham_{\rm SC} + \Ham_{\rm sub}
\end{align}
with
\begin{align}
	\Ham_{\rm SC}(t) &= -\sum_{\substack{nm \\ \alpha\beta\sigma}}  h_{n-m}^{\alpha\beta}~ e^{-i (\mathbf{r}_{n\alpha} - \mathbf{r}_{m\beta}) \cdot \mathbf{A}(t)}~ \CD{n\alpha\sigma} \C{m\beta\sigma} \notag\\
		&+ \sum_{\substack{nm\alpha\beta \\ \sigma\sigma'}} V_{n-m}^{\alpha\beta}~ \ND{n\alpha\sigma} \ND{m\beta\sigma'} - \mu \sum_{n\alpha\sigma} \ND{n\alpha\sigma} \\
	\Ham_{\textrm{sub}} &=  \sum_{\substack{n\k \\ \alpha\sigma}} \left[ g^\alpha_{n\k}~ \DD{n\k\sigma} \C{\k\alpha\sigma} + \hc \right] + \sum_{n\k \sigma} \omega_{n\k} \DD{n\k\sigma} \D{n\k\sigma}
\end{align}
where $n,m$ and $\alpha,\beta$ denote site and orbital indices, and $h_{n-m}^{\alpha\beta}$, $V_{n-m}^{\alpha\beta}$ parameterize hopping and interaction matrix elements, respectively. Light minimally couples to electrons via Peierls substitution, with $\mathbf{r}_{n\alpha}$ the real-space positions of individual sites and $\mathbf{A}(t)$ the pump field. Furthermore, $g_{n\k}^\alpha$ is the tunnel coupling to the substrate, with $\omega_{n\k}$ denoting the substrate dispersion.

The interaction term is taken to be attractive, and is treated in a mean-field approximation for the pairing instability. The corresponding self-consistent dynamics are described by a generic multi-band BCS model
\begin{align}
	\Ham = \sum_{\k} \left[\begin{array}{ll} \boldsymbol{\Psi}^\dag_{\k\uparrow} & \boldsymbol{\Psi}_{-\k\downarrow} \end{array} \right] \cdot \mathcal{H}_\k(t) \cdot \left[\begin{array}{ll} \boldsymbol{\Psi}_{\k\uparrow} \\ \boldsymbol{\Psi}^\dag_{-\k\downarrow} \end{array} \right]
\end{align}
with the BdG Hamiltonian
\begin{align}
	\mathcal{H}_\k(t) = \left[\begin{array}{ll} \mathbf{h}_\k(t) & \boldsymbol{\Delta}_\k(t) \\ \boldsymbol{\Delta}_\k^\dag(t) & -\mathbf{h}^\star_{-\k}(t) \end{array} \right]
\end{align}
where $\left[ \mathbf{h}_\k(t) \right]^{\alpha\beta} = \sum_n h_{n}^{\alpha\beta} ~e^{-i \mathbf{r}_n \k} $, and
\begin{align}
	\Delta^{\alpha\beta}_{\k} &= \frac{1}{L} \sum_{\k'} V^{\alpha\beta}_{\k-\k'} \expect{ \C{-\k',\beta\downarrow} \C{\k',\alpha\uparrow} }
\end{align}
is the multi-band order parameter, subsequently decomposed in terms of the appropriate irreducible representations of the point group as shown in the main text.

\subsubsection*{Equal-Time Keldysh Equations of Motion}

We  now derive the dissipative equal-time equations of motion for superconducting order parameters, for a multi-band BdG superconductor coupled to a fermionic bath. 
The first step is to derive dissipative equal-time equations of motion. We start from the Keldysh equations of motion in Nambu basis
\begin{align}
	i \partial_t \mathcal{G}_\k(t,t') &= \mathcal{H}_\k(t) \mathcal{G}_\k(t,t') + \int d\tau \hat{\Sigma}_\k(t,\tau) \mathcal{G}_\k(\tau,t') \\
	-i \partial_{t'} \mathcal{G}_\k(t,t') &= \mathcal{G}_\k(t,t') \mathcal{H}_\k(t') + \int d\tau \mathcal{G}_\k(t,\tau) \hat{\Sigma}_\k(\tau,t') 
\end{align}
where the Keldysh Green's function is written in Larkin-Ovchinnikov form
\begin{align}
	\mathcal{G}_\k(t,t') &= \left[\begin{array}{cc} \mathcal{G}_\k^R(t,t') & \mathcal{G}_\k^K(t,t') \\ & \mathcal{G}_\k^A(t,t') \end{array}\right]
\end{align}
The self energy contains the dissipative coupling to the electronic reservoir, and reads explicitly
\begin{align}
	\hat{\Sigma}_\k(t,t') &= \iint \frac{d\Omega d\omega}{2\pi}~ e^{-i\Omega(t-t')} \left| g_{n\k} \right|^2 \rho_\k(\omega) ~\times \notag\\
	& ~~~~~\times \left[\begin{array}{cc}  \frac{1}{\Omega - \omega + i\eta} & -\frac{2i\eta \tanh\left( \frac{\beta \Omega}{2} \right)}{(\Omega - \omega)^2 + \eta^2} \\ & \frac{1}{\Omega - \omega - i\eta}  \end{array}\right] 
\end{align}
where $\rho_\k(\omega)$ is the transverse partial density of states for a 3D bulk/substrate.

In the wide-band limit, $\rho_\k(\omega) \to \rho_0$ is taken to be constant. This amounts to a Markov approximation of the retarded component, and the self energy becomes
\begin{align}
	\Sigma_\k(t,t') &= \int \frac{d\Omega}{2\pi}~ e^{-i\Omega(t-t')} \left[\begin{array}{cc} -i\Gamma & -2i\Gamma \tanh\left(\frac{\beta \Omega}{2}\right) \\ & i\Gamma \end{array}\right] \\
	&= \left[\begin{array}{cc}  -i\Gamma~ \delta(t-t') & -2\Gamma \left[ \beta  \sinh \left( \frac{\pi (t-t')}{\beta} \right) \right]^{-1} \\ & i\Gamma~ \delta(t-t') \end{array}\right]
\end{align}
where $\Gamma = \pi \rho_0 |g|^2$ denotes the relaxation constant. This approximation to the fermionic self-energy is unnecessary in principle, but significantly expedites the numerical simulation. In practice, a microscopic description of the system bath coupling and a finite substrate band width will alter the dissipative dynamics on short time scales. However, since the primary role of the bath in this work is to permit the system to dissipate energy absorbed from the pump and return to thermal equilibrium at long times, such details of the dissipative coupling are irrelevant to the conclusions presented in the main text.

The next task is to derive an equal-time equation of motion for the order parameters. Starting from the Kadanoff-Baym equation, the evolution of the retarded Green's function becomes
\begin{align}
	i \partial_t \mathcal{G}_\k^R(t,t') &= \mathcal{H}_\k(t) \mathcal{G}_\k^R(t,t') + \int_{t'}^{t} d\tau~ \Sigma^R(t,\tau) \mathcal{G}_\k^R(\tau,t') \\
		&= \left[\mathcal{H}_\k(t) - i\Gamma \right] \mathcal{G}_\k^R(t,t')
\end{align}
with $\mathcal{G}^A(t,t') = \left[ \mathcal{G}^R(t',t) \right]^\dag$. The evolution of the Keldysh Green's function reads
\begin{align}
	\partial_{t} \mathcal{G}_\k^K(t,t') &= -i \left[ \mathcal{H}_\k(t) - i \Gamma \right] \mathcal{G}_\k^K(t,t') \notag\\
		&- i \int_{-\infty}^{t'} d\tau~ \Sigma^{K}(t,\tau) \mathcal{G}_\k^A(\tau,t') \\
	\partial_{t'} \mathcal{G}_\k^K(t,t') &= +i \mathcal{G}_\k^K(t,t') \left[ \mathcal{H}_\k(t') + i \Gamma \right] \notag\\
		&+ i \int_{-\infty}^{t} d\tau~ \mathcal{G}_\k^R(t,\tau) \Sigma^K(\tau,t')
\end{align}
Define now the equal-time expectation values
\begin{align}
	N_\k^{\alpha\beta}(T) = \expect{ \CD{\k\beta} (T) \C{\k\alpha} (T) } = -\frac{i}{2} \mathcal{G}^K_{\k,\alpha\beta}(T, T) + \frac{1}{2}
\end{align}
Its equation of motion can be readily derived as
\begin{align}
	\partial_T N_\k(T) &= -\frac{i}{2} \left. \left( \partial_t + \partial_{t'} \right) \mathcal{G}_\k^K(t,t') \right|_{t=t'=T} \\
		&= -i \left[  \mathcal{H}_\k(T),~ N_\k(T)  \right] - \Gamma \left\{ 2 N_\k(T) - 1 - I_\k^K(T) \right\} 
\end{align}
with
\begin{align}
	I_\k^{K}(T) &= \frac{1}{2\Gamma} \int_{-\infty}^{T} d\tau \left[ \mathcal{G}_\k^R(T,\tau) + \mathcal{G}_\k^R(T,\tau)^\dag \right] \Sigma^K(\tau-T) \notag\\
		= \frac{1}{2\Gamma}& \int_{0}^{\infty} d\tau \left[ \mathcal{G}_\k^R(T,T-\tau) + \mathcal{G}_\k^R(T,T-\tau)^\dag \right] \Sigma^K(\tau)
\end{align}
with
\begin{align}
	\Sigma^K(t-t') = 2\Gamma \left[ \beta \sinh \left( \frac{\pi (t-t')}{\beta} \right) \right]^{-1}
\end{align}
and
\begin{align}
	\partial_T \mathcal{G}_\k^R(T,T-\tau) &= i \mathcal{G}_\k^R(T,T-\tau) \left[ \mathcal{H}_\k(T-\tau) + i\Gamma \right] \notag\\
		&- i \left[ \mathcal{H}_\k(T) - i\Gamma \right] \mathcal{G}_\k^R(T,T-\tau)
\end{align}

\subsection*{Dissipative BdG Equations of Motion}

Now, define the retarded Nambu Green's functions
\begin{align}
	\mathcal{G}_\k^R(t,t') = \left[\begin{array}{ll} \mathbf{G}^R_{\k} & \mathbf{D}^R_{\k} \\ -\nu \left( \mathbf{D}^R_{-\k} \right)^\star &  -\left( \mathbf{G}^R_{-\k} \right)^\star \end{array}\right]
\end{align}
where $\nu = +1$ for singlet pairing, $\nu = -1$ for triplet pairing,
with normal and anomalous multiband components
\begin{align}
	G^R_{\k,\alpha\beta} &= -i \theta(t-t') \expect{ \left\{ \CD{\k\alpha\sigma}(t),~ \C{\k\beta\sigma}(t') \right\} } \\
	D^R_{\k,\alpha\beta} &= -i \theta(t-t') \expect{ \left\{ \CD{\k\alpha\uparrow}(t),~ \CD{-\k\beta\downarrow}(t') \right\} }
\end{align}
Furthermore, define the equal-time function
\begin{align}
	\mathbf{N}_{\k}(t) = \left[\begin{array}{ll} \mathbf{n}_{\k}(t) & \mathbf{f}_{\k}(t) \\ \mathbf{f}_{\k}^\dag(t) & 1 - \mathbf{n}^\star_{-\k}(t) \end{array}\right]
\end{align}
with expectation values
\begin{align}
	n_{\k,\alpha\beta}(t) &= \expect{ \CD{\k\beta\sigma} \C{\k\alpha\sigma} (t) } \\
	f_{\k,\alpha\beta}(t) &= \expect{ \CD{\k\beta\uparrow} \CD{-\k\alpha\downarrow} (t) }
\end{align}
Substituting into the equal-time equations of motion derived above, we finally arrive at the closed set of non-linear equations of motion
\begin{align}
	i \partial_t \mathbf{n}_\k(t) &= \left[ \mathbf{h}_\k(t),~  \mathbf{n}_\k(t) \right] + \boldsymbol{\Delta}_\k(t)~ \mathbf{f}^\dag_{\k}(t) - \mathbf{f}_{\k}(t)~ \boldsymbol{\Delta}^\dag_\k(t) \notag\\
		&- \Gamma \left( 2 \mathbf{n}_\k(t)  - 1 - \mathbf{I}^G_\k(t) \right) \label{eq:FirstEoM} \\
	i \partial_t \mathbf{f}_\k(t) &= \mathbf{h}_\k(t) \mathbf{f}_\k(t) + \mathbf{f}_\k(t) \mathbf{h}_{-\k}(t)  \notag\\
		&- \boldsymbol{\Delta}_k(t) \left( \mathbf{n}^\star_{-\k}(t) - 1 \right) - \mathbf{n}_\k(t) \boldsymbol{\Delta}_k(t) \notag\\
		&- \Gamma \left( 2 \mathbf{f}_\k(t) - \mathbf{I}^D_\k(t) \right) \\
	\boldsymbol{\Delta}_k(t) &= \sum_{\k'} \hat{V}_{\k-\k'}~ \mathbf{f}_{\k}^\star(t) \\
	\mathbf{I}^G_\k(t) &= \frac{1}{2\Gamma} \int_0^\infty d\tau \left[ \mathbf{G}^R_\k(t,t-\tau) + (\mathbf{G}^R_\k(t,t-\tau))^\dag \right] \Sigma^K(\tau) \\
	\mathbf{I}^D_\k(t) &= \frac{1}{2\Gamma} \int_0^\infty d\tau \left[ \mathbf{D}^R_\k(t,t-\tau) + (\mathbf{D}^R_\k(t,t-\tau))^\dag \right] \Sigma^K(\tau) \\
	i \partial_t \mathbf{G}^R_\k&(t,t') = \left( \mathbf{h}_\k(t) - i \Gamma \right) \mathbf{G}^R_\k(t,t') - \nu \boldsymbol{\Delta}_\k(t) \left( \mathbf{D}^R_\k(t,t')\right)^\star \\
	i \partial_t \mathbf{D}^R_\k&(t,t') = \left( \mathbf{h}_\k(t) - i \Gamma \right) \mathbf{D}^R_\k(t,t') - \boldsymbol{\Delta}_\k(t) \mathbf{G}^R_\k(t,t') \label{eq:LastEoM}
\end{align}
All simulations presented in the main text follow from numerically integrating the integro-differential equations of motion (\ref{eq:FirstEoM})-(\ref{eq:LastEoM}), after decomposing $\hat{V}_\k$ and $\boldsymbol{\Delta}_\k$ into the appropriate irreducible representations. We stress that while the wide-band approximation for the fermionic bath permits omitting a calculation of two-time Keldysh Green's functions $G_\k^K(t,t')$, $D_\k^K(t,t')$ in favor of their equal-time values only, the equations nevertheless necessitate solving for the full two-time dependence of the retarded normal and anomalous Green's functions $G_\k^R(t,t')$, $D_\k^R(t,t')$, which in turn depend on the equal-time magnitude of the superconducting gap function.

\end{document}